\documentclass[twocolumn, longauthor]{aastex631}

\usepackage{xcolor}
\usepackage{comment}
\usepackage{graphicx}
\usepackage{dcolumn}
\usepackage{bm}
\usepackage{hyperref}
\DeclareUnicodeCharacter{2009}{\,}

\begin{document}

\title{Analytic understanding of the resonant nature of Kozai Lidov Cycles
 with a precessing quadrupole potential}

\correspondingauthor{Ygal Y. Klein}
\email{ygalklein@gmail.com}

\author[0009-0004-1914-5821]{Ygal Y. Klein}
\affiliation{Dept. of Particle Phys. \& Astrophys., Weizmann Institute of Science,
 Rehovot 76100, Israel}

\author[0000-0003-0584-2920
]{Boaz Katz}
\affiliation{Dept. of Particle Phys. \& Astrophys., Weizmann Institute of Science,
 Rehovot 76100, Israel}







\begin{abstract}

 The very long-term evolution of the hierarchical restricted three-body problem with a slightly aligned precessing quadrupole potential is studied analytically. This problem describes the evolution of a star and a planet which are perturbed either by a (circular and not too inclined) binary star system or by one other star and a second more distant star, as well as a perturbation by one distant star and the host galaxy or a compact-object binary system orbiting a massive black hole in non-spherical nuclear star clusters \citep{hamers2017,petrovich2017}. Previous numerical experiments have shown that when the precession frequency is comparable to the Kozai-Lidov time scale, long term evolution emerges that involves extremely high eccentricities with potential applications for a broad scope of astrophysical phenomena including systems with merging black holes, neutron stars or white dwarfs. By averaging the secular equations of motion over the Kozai-Lidov Cycles (KLCs) we solve the problem analytically in the neighborhood of the KLC fixed point where the eccentricity vector is close to unity and aligned with the quadrupole axis and for a precession rate similar to the Kozai Lidov time scale. In this regime the dynamics is dominated by a resonance between the perturbation frequency and the precession frequency of the eccentricity vector. While the quantitative evolution of the system is not reproduced by the solution far away from this fixed point, it sheds light on the qualitative behaviour.

\end{abstract}



\section{Introduction} \label{sec:intro}

In this letter we study analytically the dynamics of a test particle
orbiting a central mass $M$ on a Keplerian orbit with semimajor
axis $a$ which is perturbed by an external
quadrupole potential given by: 
\begin{equation}
\Phi_{outer}=\frac{\Phi_0}{a^2}\left[3\left(\mathbf{\hat{j}}_{outer}\cdot\mathbf{r}\right)^{2}-r^{2}\right]\label{eq:potential}
\end{equation}
where $\Phi_0$ is constant. In the periodic analytically solved
Kozai-Lidov cycles (KLCs) \citep{lidov1962,kozai62} the external quadrupole potential is constant in time (i.e 
 $\mathbf{\hat{j}}_{outer}$ is a constant unit vector) (for a recent review on KLCs see \citep{naoz2016}). We study the case where the quadrupole potential is time dependent and $\mathbf{\hat{j}}_{outer}$ is a unit vector which precesses around the $z$ axis at a constant rate $\beta$ with a constant inclination $\alpha$:
\begin{equation}
 \mathbf{\hat{j}}_{outer}=\left(\begin{array}{c}
   \sin\alpha\cos\left(\beta\tau\right) \\
   -\sin\alpha\sin\left(\beta\tau\right) \\   
   \cos\alpha
  \end{array}\right)\label{eq:jOuter_as_a_function_of_tau}
\end{equation}
where $\tau\equiv\frac{t}{t_{sec}}$ and $t_{sec}=\frac{\sqrt{GMa}}{\Phi_0}$ is the secular timescale.

This problem describes the evolution of a star and a planet which are perturbed either by a (circular and not too inclined) binary star system or by one other star and a second more distant star \citep{hamers2017}, as well as a perturbation by one distant star and the host galaxy or a compact-object binary system orbiting a massive black hole in non-spherical nuclear star clusters \citep{petrovich2017}. Previous numerical experiments have shown that when the precession frequency is comparable to the Kozai-Lidov time scale, long term evolution emerges that involves extremely high eccentricities \citep{hamers2017} with potential applications for the formation of planets around white dwarfs \citep{munoz20,oconnor21,stephan21} and hot planets \citep{fabrycky2007,katz2011,naoz2011,grishin18}. If the test particle assumption is relaxed, the system exhibits similar dynamics and the description is applicable to a broader scope of astrophysical phenomena, including Type Ia supernovae through the merger or collision of white dwarfs in multiple systems \citep{thompson2011,katz2012,pejcha2013,fang2018,grishin22}, gravitational wave emission through the merger of black holes or neutron stars in quadruple systems \citep{liu2019,safarzadeh2020,hamers2020} and the formation of close binaries \citep{antonini2012,petrovich2017,bub2020,grishin22}.

As mentioned, the case of $\alpha=0$ is the periodic analytically solved
Kozai-Lidov cycles (KLCs) \citep{lidov1962,kozai62}.

\section{Equations of motion} \label{sec:equations_of_motion}

\begin{figure}
 \begin{centering}
 \includegraphics[scale=0.21]{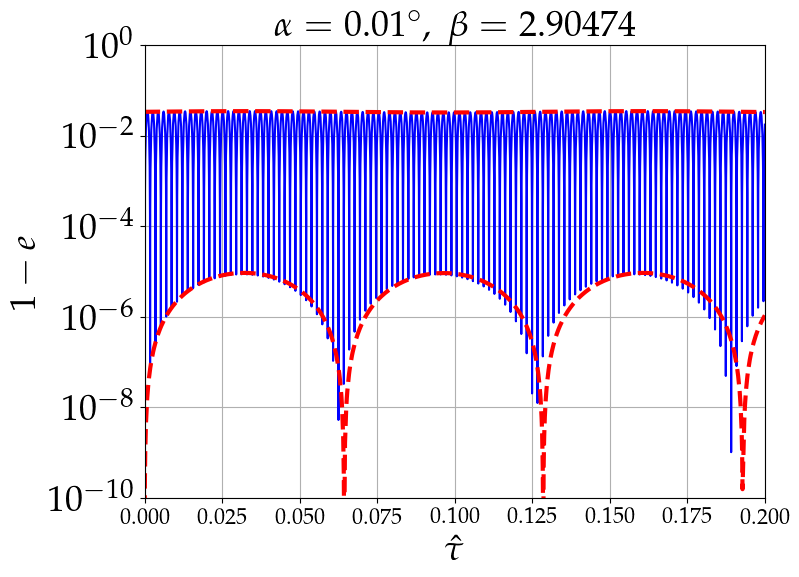}\includegraphics[scale=0.21]{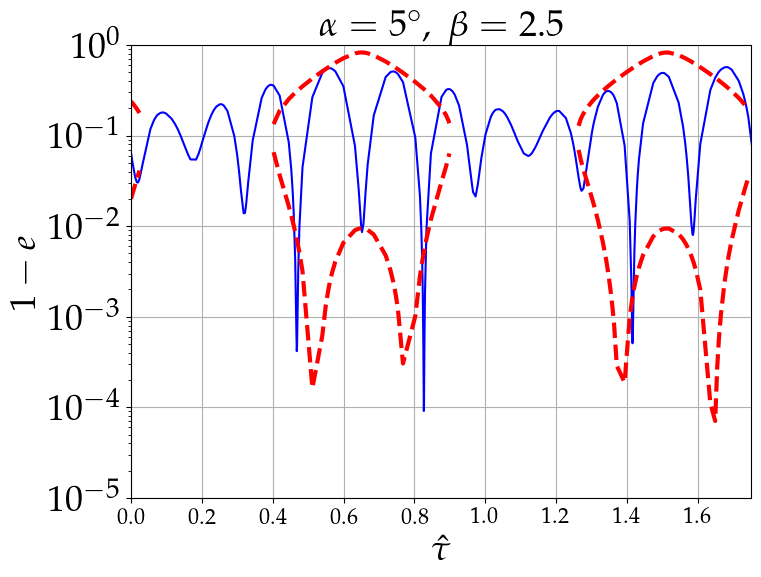}
  \par\end{centering}
 \begin{centering}
  \includegraphics[scale=0.2]{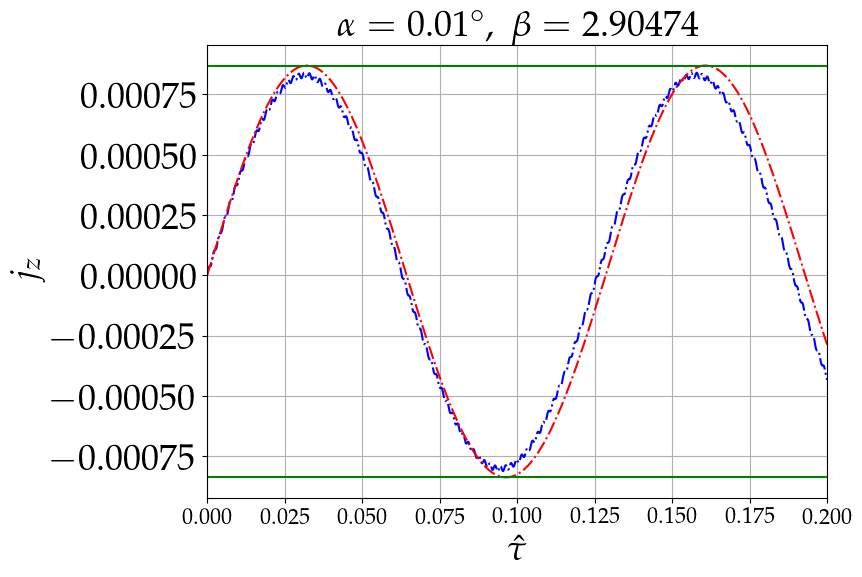}\includegraphics[scale=0.2]{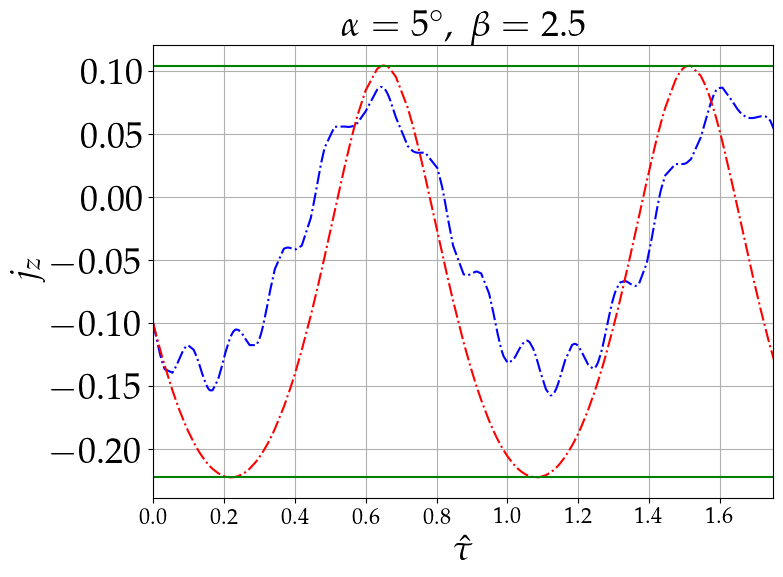}
  \par\end{centering}
 \begin{centering}
  \includegraphics[scale=0.21]{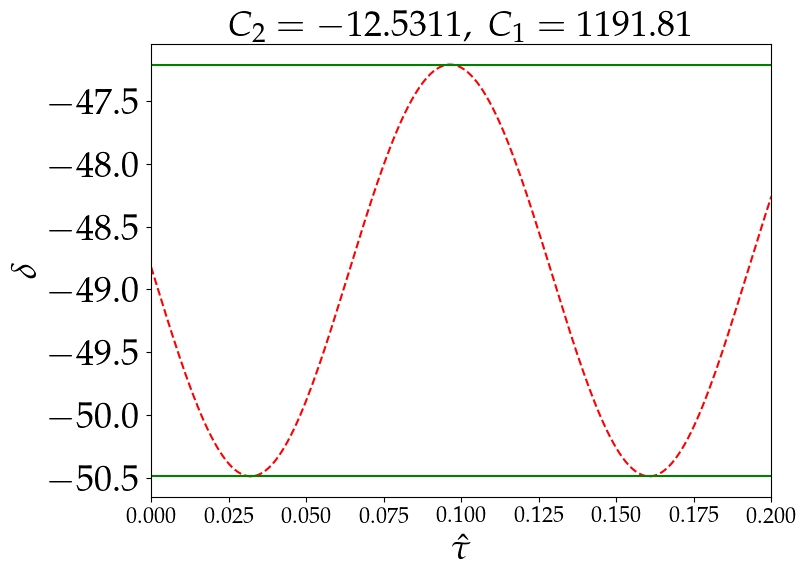}\includegraphics[scale=0.21]{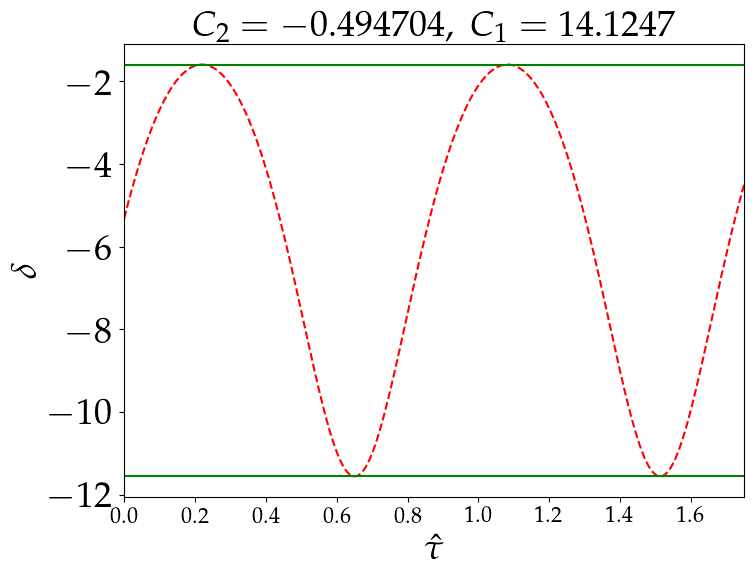}
  \par\end{centering}
 \caption{Results of numerical integrations for: Left panel: $\alpha=0.01^{\circ}$ and $\beta\approx2.9$,
 with initial conditions $e_{x}=j_{x}=-j_{y}=10^{-5},e_{z}=0.98$ and right panel: $\alpha=5^{\circ}$ and $\beta=2.5$,
 with initial conditions $e_{x}\sim-10^{-5}, j_{x}\sim-0.28,j_{y}\sim-0.186,e_{z}\sim0.825$.
 The blue solid lines are the result of the integration of the full
 secular equations, Eqs. \ref{eq:secular_equations} (with \ref{eq:jOuter_as_a_function_of_tau}),
 while the red dashed lines are the result of the averaged equations,
 Eqs. \ref{eq:splusDot}-\ref{eq:a_plus_b_dot}, and using Eq. \ref{eq:polynom_for_extremal_eccentricity} (where it has real roots) to determine $e_{min}$ and $e_{max}$  using Eqs. \ref{eq:C4}
 and \ref{eq:Ck}. The two green horizontal lines in the bottom panel
 represent the extremum values of $\delta$ as determined from initial
 conditions and in the middle panel the maximal and minimal values
 of $j_{z}$ as determined from initial conditions using Eq. \ref{eq:C4}
 and the extremums of $\delta$. $\hat{\tau}$ is defined in Eq. \ref{eq:tau_hat}.\label{fig:delta_and_jz_one_minus_e_alpha_0.01_ez_0.98}}
\end{figure}

The dynamics of the test particle can be parameterized
by two dimensionless orthogonal vectors $\mathbf{j}=\mathbf{J}/\sqrt{GMa}$, where $\mathbf{J}$ is the specific angular momentum vector, and
$\mathbf{e}$ a vector pointing in the direction of the pericenter
with magnitude $e$. In the secular approximation, $a$ is constant with time while $\mathbf{j}$ and
$\mathbf{e}$ evolve according to the Kozai-Lidov equations (as Eq. 10a-b in \citep{hamers2017})
\begin{eqnarray}
\frac{d\mathbf{j}}{d\tau}=&\frac{3}{4}\left(\left(\mathbf{j}\cdot\mathbf{\hat{j}}_{outer}\right)\mathbf{j}-5\left(\mathbf{e}\cdot\mathbf{\hat{j}}_{outer}\right)\mathbf{e}\right)\times\mathbf{\hat{j}}_{outer} \cr
\frac{d\mathbf{e}}{d\tau}=&\frac{3}{2}\left(\mathbf{j}\times\mathbf{e}\right)-\frac{3}{4}\left(5\left(\mathbf{e}\cdot\mathbf{\hat{j}}_{outer}\right)\mathbf{j}-\left(\mathbf{j}\cdot\mathbf{\hat{j}}_{outer}\right)\mathbf{e}\right)\times\mathbf{\hat{j}}_{outer} \cr
 \label{eq:secular_equations}
\end{eqnarray}
with $\mathbf{\hat{j}}_{outer}$ given by Eq. \ref{eq:jOuter_as_a_function_of_tau} (itself a solution of Eq. 10c in \citep{hamers2017}). A numerical integration of Eqs. \ref{eq:secular_equations} is shown as blue lines in the top two panels of Fig. \ref{fig:delta_and_jz_one_minus_e_alpha_0.01_ez_0.98} for $\alpha=0.01^{\circ}$ and $\beta\approx2.9$ (left panel) and $\alpha=5^{\circ}$ and $\beta=2.5$ (right panel). We remind that in the $\alpha=0$ case (pure KLCs), $j_z$ (middle panel of Fig. \ref{fig:delta_and_jz_one_minus_e_alpha_0.01_ez_0.98}) is constant and $e$ (top panel of Fig. \ref{fig:delta_and_jz_one_minus_e_alpha_0.01_ez_0.98}) is periodically oscillating but with a constant $e_{max}$ (which for $e_0\ll1$ can be approximated with $e_{max}\approx\sqrt{1-\frac{5}{3}j^2_z}$). As can be seen in the top and middle panels of Fig.  \ref{fig:delta_and_jz_one_minus_e_alpha_0.01_ez_0.98} the times of zero crossing of $j_z$ correspond to the times of extremely high eccentricities, as expected from KLCs.

We restrict the analysis to the regime where $\alpha\ll1$ (i.e $\alpha$ being a
small parameter around which $\alpha=0$ is already analytically solved)
and $\left|\mathbf{e}\cdot\mathbf{\hat{j}}_{outer}\right|\sim1$ (i.e
$\mathbf{j}\cdot\mathbf{\hat{j}}_{outer} \approx j_z \ll1$, $e\sim1$ and inclination close to $90^\circ$, which is close to the KLC fixed point of $e=1$, $i=90^\circ$ and $j=0$). In this regime, the eccentricity vector precesses around the $z$ axis. When the frequencies of the precession of $\mathbf{e}$ and $\mathbf{\hat{j}}_{outer}$ are far from each other - the precession of the quadrupole potential has a minor effect on the KLCs. On the other hand, when these two frequencies are close, long-term resonant dynamics are obtained and are the focus of this letter.

\section{Approximated Equations} \label{sec:Approximated Equations}

In this regime and up to first order in $\alpha$ one obtains the
following 6 equations (neglecting $j_{z}$ in this regime in the rhs of the
derivatives of $e_x$, $e_y$ and $e_z$):
\begin{eqnarray}
 \frac{d}{d\tau}j_{z}&=\frac{15}{4}e_{z}\alpha\left(e_{x}\sin\left(\beta\tau\right)+e_{y}\cos\left(\beta\tau\right)\right)\label{eq:jzdot} \\
 \frac{d}{d\tau}e_{z}&=\frac{3}{4}\left(2\left(j_{x}e_{y}-j_{y}e_{x}\right)+5e_{z}\alpha\left(j_{x}\sin\left(\beta\tau\right)+j_{y}\cos\left(\beta\tau\right)\right)\right)\label{eq:ezdot}
\end{eqnarray}
\begin{eqnarray}
 \frac{d}{d\tau}e_{x} & =-\frac{9}{4}e_{z}j_{y}\label{eq:exdot} \\
 \frac{d}{d\tau}e_{y} & =+\frac{9}{4}e_{z}j_{x}\label{eq:eydot}
\end{eqnarray}
\begin{eqnarray}
 \frac{d}{d\tau}j_{x} & =-\frac{15}{4}e_{z}\left(e_{y}+e_{z}\alpha\sin\left(\beta\tau\right)\right)\label{eq:jxdot} \\
 \frac{d}{d\tau}j_{y} & =+\frac{15}{4}e_{z}\left(e_{x}-e_{z}\alpha\cos\left(\beta\tau\right)\right)\label{eq:jydot}
\end{eqnarray}
In the lowest order approximation,
$\frac{d}{d\tau}e_{z}=0$, resulting with a forced harmonic
oscillator for the vector $\mathbf{e}$ in the $x-y$ plane with $\ddot{e}_{x}=\omega_{0}^{2}\left(L\cos\left(\omega\tau\right)-e_{x}\right)$
where $\omega=\beta,L=e_{z}\alpha$ and $\omega_{0}=\sqrt{\frac{135}{16}}e_{z}$.
Below we solve the next level of approximation where $e_{z}$ is slowly changing.

\section{Averaged Equations\label{sec:Averaged Equations}}

Since $\alpha$ is small the dynamics on short time scales follow
the known (test particle triple system) Kozai-Lidov Cycles, which
have two constants of motion: $j_{z}$ and
\begin{equation}
C_{K}=e^{2}-\frac{5}{2}e_{z}^{2}=e^{2}\left(1-\frac{5}{2}\sin^{2}i\sin^{2}\omega\right).
\end{equation}
On longer time scales the parameters of the KLC, $j_{z}$ and $C_{K}$,
evolve.

Consider the following ansatz for the vector $\mathbf{e}$
in the $x-y$ plane: At any time $\tau$, the projection of the vector
$\mathbf{e}$ on the $x-y$ plane can be presented as a point moving
on a slowly evolving ellipse with semimajor axis $a$ inclined with an angle $\theta$ with respect to the $x$ axis and semiminor axis $b$ centered at the origin, i.e
\begin{equation}
 \mathbf{e}_{x-y}=\alpha^{\frac{1}{3}}\left(\begin{array}{cc}
   \cos\theta, & -\sin\theta \\
   \sin\theta, & \cos\theta
  \end{array}\right)\left(\begin{array}{c}
   a\cos\left(\hat{\beta}\hat{\tau}+\phi\right) \\
   b\sin\left(\hat{\beta}\hat{\tau}+\phi\right)
  \end{array}\right)\label{eq:exy-ansatz}
\end{equation}
where $\phi$ is a slowly dynamically evolving phase and
\begin{equation}
 \hat{\tau}=\frac{1}{2}\alpha^{\frac{2}{3}}\tau
 \label{eq:tau_hat}
\end{equation}
and
\begin{equation}
 \hat{\beta}=2\alpha^{-\frac{2}{3}}\beta.
 \label{eq:beta_hat}
\end{equation}
See note after Eq. \ref{eq:a_plus_b_dot} regarding the choice of normalization prefactors: $\alpha^{\frac{1}{3}},\alpha^{\frac{2}{3}}$ and $\alpha^{-\frac{2}{3}}$.
The ansatz in Eq. \ref{eq:exy-ansatz} has a symmetry under the following transformation (both changes together)
\begin{eqnarray}
\left(a-b\right)\rightarrow-\left(a-b\right)\nonumber\\
\left(\theta-\phi\right)\rightarrow\left(\theta-\phi+\pi\right)\nonumber
\end{eqnarray}
meaning that without loss of generality $\left(a-b\right)$ is non negative.

Using Eqs. \ref{eq:exdot}-\ref{eq:eydot} in the limit $e_{z}=1$ and neglecting the time derivatives of the slowly varying functions, the projection of the angular momentum on the $x-y$ plane is correspondingly given by
\begin{equation}
 \mathbf{j}_{x-y}=\frac{4}{9}\alpha^{\frac{1}{3}}\beta\left(\begin{array}{cc}
   \cos\theta, & -\sin\theta \\
   \sin\theta, & \cos\theta
  \end{array}\right)\left(\begin{array}{c}
   b\cos\left(\hat{\beta}\hat{\tau}+\phi\right) \\
   a\sin\left(\hat{\beta}\hat{\tau}+\phi\right)
  \end{array}\right).\label{eq:jxy}
\end{equation}
Note the ansatz includes four slowly evolving variables, $a,b,\theta,\phi$, which describe the averaged evolution of the four components $e_x,e_y,j_x,j_y$.

Since the frequency of the precession of $\mathbf{\hat{j}}_{outer}$ is $\beta$ and the driving frequency of the Kozai oscillations is $\sqrt{\frac{135}{16}}e_{z}$ a resonance is obtained between the two perturbations when the two frequencies approach each other and it is useful to quantify the distance from resonance by a dynamical parameter,
\begin{equation}
 \delta=\alpha^{-\frac{2}{3}}\frac{1}{\beta_{0}}\left(\left(\beta_{0}\bar{e}_{z}\right)^{2}-\beta^{2}\right)\label{eq:delta}
\end{equation}
where $\bar{e}_{z}$ is the averaged value of $e_{z}$ over KLC which satisfies (using Eq. \ref{eq:ezdot})
\begin{equation}
 e_{z}=\bar{e}_{z}+\frac{\alpha^{\frac{2}{3}}}{6}\left(a^{2}-b^{2}\right)\cos\left(2\left(\hat{\beta}\hat{\tau}+\phi\right)\right)\label{eq:instantanous_ez}
\end{equation}
and
\begin{equation}
 \beta_{0}=\sqrt{\frac{135}{16}}\approx2.9\label{eq:beta0}.
\end{equation}
Using the following slow variables 
\begin{eqnarray}
 s=-\frac{45}{2}\left(a-b\right)\sin\left(\theta-\phi\right)\\
 c=-\frac{45}{2}\left(a-b\right)\cos\left(\theta-\phi\right)
\end{eqnarray}
and focusing on the resonant limit of $\omega=\omega_{0}$ in the forced harmonic oscillator mentioned above, i.e $\beta=\beta_{0}$, the following set of ODEs is obtained:
\begin{eqnarray}
 \dot{\delta}=&s\label{eq:deltaDot}\\
 \dot{s}=&-\left(45\beta_{0}+\delta c\right)\label{eq:splusDot}\\
 \dot{c}=&\delta s\label{eq:cPlusDot}
\end{eqnarray}
and 
\begin{eqnarray}
 \frac{d\left(\theta+\phi\right)}{d\hat{\tau}} & =\delta\label{eq:theta_minus_phi_dot} \\
 \frac{d\left(a+b\right)}{d\hat{\tau}}     & =0\label{eq:a_plus_b_dot},
\end{eqnarray}
where $\dot{}$ denotes a derivative with respect to $\hat{\tau}$.

Several notes are in order: (1) The parameters $s,c,a+b,\theta+\phi$ uniquely determine all the slow variables: $a,b,\theta,\phi$. (2) The evolution of $\delta,s$ and $c$ can be obtained by solving the closed subset of Eqs. \ref{eq:deltaDot}-\ref{eq:cPlusDot}. (3) The equations obtained have no explicit dependence on the small parameter $\alpha$. In fact, the $\alpha$ dependent prefactors in Eqs. \ref{eq:exy-ansatz}-\ref{eq:delta} were chosen for this reason. (4) $j_{z}$ can be obtained from the demand that $\mathbf{j}\cdot\mathbf{e}=0$. In fact, the following combination of $j_z$ and $\delta$ is constant:
\begin{equation}
j_z+\frac{\delta}{6}\alpha^{\frac{2}{3}}=\text{const.},\label{eq:C4}
\end{equation}
allowing $j_z$ to be readily obtained using the initial conditions and the time evolution of $\delta$.
The resulting slow evolution of $\delta$ for the examples in Fig. \ref{fig:delta_and_jz_one_minus_e_alpha_0.01_ez_0.98} is shown in the bottom panel and is used for the solution of $j_z$ plotted as a dashed red line in the middle panel. As can be seen in the left panel, the slow evolution of $j_z$ agrees to an excellent approximation with the numerical result.

\section{Analytic Solution\label{sec:Analytic-Solution}}

The averaged equations, Eqs. \ref{eq:deltaDot}-\ref{eq:cPlusDot}, admit two constants of the motion, denoted $C_{1}$ and $C_{2}$,
\begin{eqnarray}
 C_{1}=&\frac{1}{2}\left(\delta^{2}+45\left(a-b\right)\cos\left(\theta-\phi\right)\right)\label{eq:C1}\\
 C_{2}=&\left(a-b\right)^{2}+\frac{2\delta}{\sqrt{15}}\label{eq:C2}
\end{eqnarray}
implying that the evolution of the three variables $\delta,a-b$ and $\theta-\phi$ is periodic.
Using Eq. \ref{eq:a_plus_b_dot} we define a third constant
\begin{equation}
 C_{3}=\left(a+b\right)^{2}\label{eq:C3}
\end{equation}
which together with $C_1$ and $C_2$ determine the long term
evolution of the entire system.

The resulting evolution of $\delta$ is equivalent to the dynamics of a particle moving in a one dimensional potential with a constant energy 
\begin{equation}
 E=\frac{1}{2}\dot{\delta}^{2}+V=\frac{1}{2}\left(\left(\frac{45}{2}\right)^{2}C_{2}-C_{1}^{2}\right)\label{eq:energy}
\end{equation}
where (using $\beta_{0}=\sqrt{\frac{135}{16}}$,
see Eq. \ref{eq:beta0})
\begin{equation}
 V=45\beta_{0}\delta-\frac{1}{2}C_{1}\delta^{2}+\frac{1}{8}\delta^{4}\label{eq:Potential}.
\end{equation}

\begin{figure}
 \begin{centering}
  \includegraphics[scale=0.22]{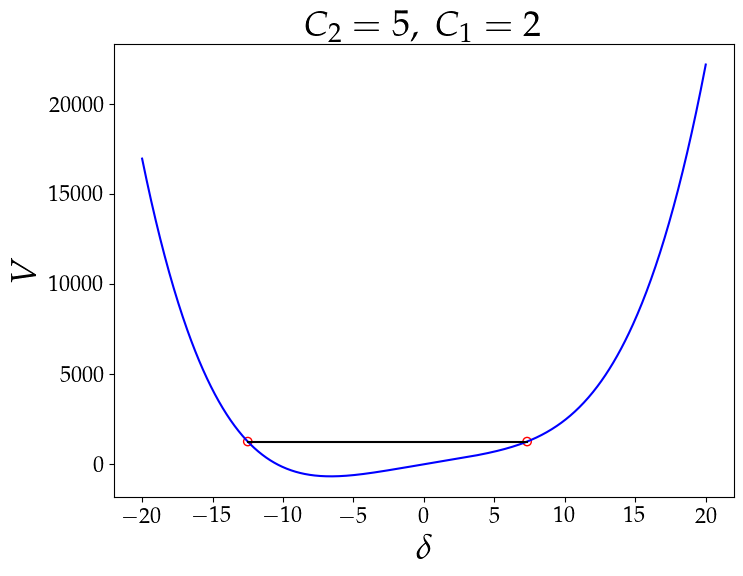}\includegraphics[scale=0.22]{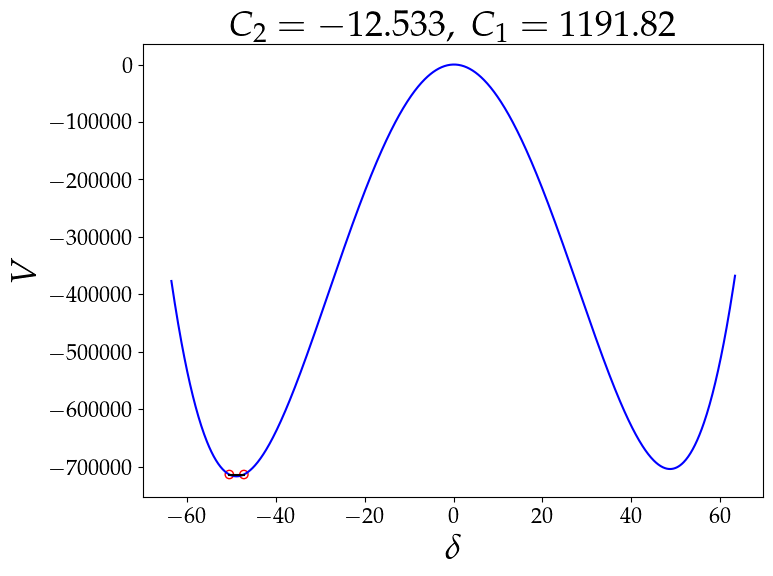}
  \par\end{centering}
 \begin{centering}
  \includegraphics[scale=0.22]{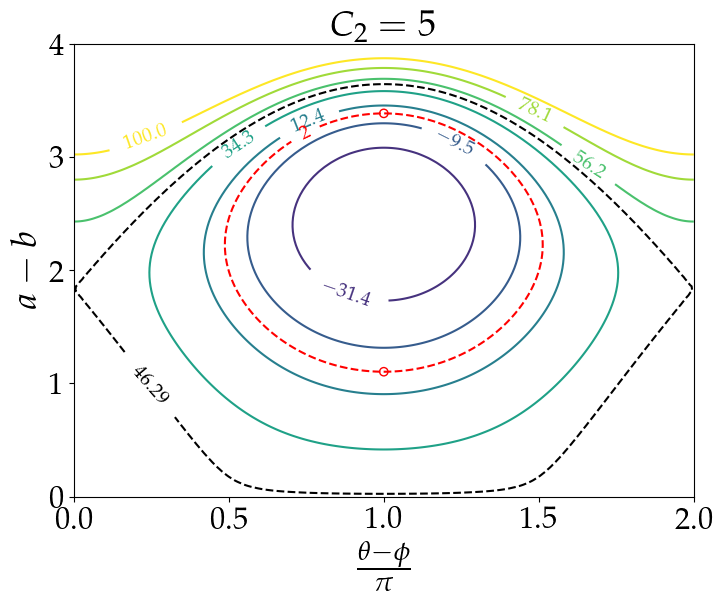}\includegraphics[scale=0.22]{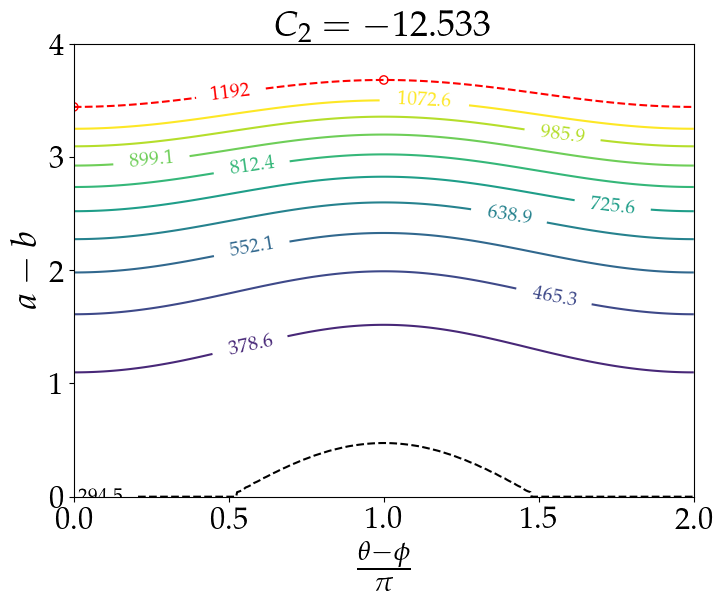}
  \par\end{centering}
 \caption{Upper panel: the potential $V$ (Eq. \ref{eq:Potential}) in blue
 and the constant energy $E$ (Eq. \ref{eq:energy}) in black for the
 two optional shapes dependent on the constants $C_{1}$ (Eq. \ref{eq:C1})
 and $C_{2}$ (Eq. \ref{eq:C2}). Lower panel: Trajectories in the
 $a-b$ vs. $\theta-\phi$ plane for different values of $C_{1}$ at
 some $C_{2}$. Dashed black line mark the minimal value of $C_{1}$
 for rotations. Red dashed lines mark the value of $C_{1}$ of the
 potentials in the upper panels. The left plots show a case where $C_{1}<C_{1}^{\text{crit}}$
 and $\theta-\phi$ is librating. The right plots are the case that
 is shown in the left panel of Fig. \ref{fig:delta_and_jz_one_minus_e_alpha_0.01_ez_0.98}
 and show a case where $C_{1}>C_{1}^{\text{crit}}$ and $\theta-\phi$
 is rotating. Red circles (in all panels) mark the extremums of $\delta$
 (which are also $a-b$ extremums, see Eq. \ref{eq:C2}).\label{fig:potential_energy_and_libration_rotation_maps}}
\end{figure}

This potential has two distinct shapes depending on whether $C_1$ is smaller or larger than the critical value
\begin{equation}
 C_{1}^{\text{crit}}=15\left(\frac{3}{2}\right)^{\frac{7}{3}}.
\end{equation}
If $C_{1}<C_{1}^{\text{crit}}$ the potential has no maxima and one
minimum (see example in the left upper panel of Fig. \ref{fig:potential_energy_and_libration_rotation_maps}).
If $C_{1}>C_{1}^{\text{crit}}$ the potential has a maxima and two minima (see example in the right upper panel of Fig. \ref{fig:potential_energy_and_libration_rotation_maps} showing the case that is solved in the left panel of Fig. \ref{fig:delta_and_jz_one_minus_e_alpha_0.01_ez_0.98}) \footnote{For analysis of the different features we equip the reader with a link to a visualization of Eqs. \ref{eq:Potential}, \ref{eq:energy}: \url{https://www.desmos.com/calculator/ubgicqtddn}}. The extremum values of $\delta$ determined from the potential and energy are marked as red circles in the top panels of Fig. \ref{fig:potential_energy_and_libration_rotation_maps} and are plotted as green lines in the bottom panel of Fig. \ref{fig:delta_and_jz_one_minus_e_alpha_0.01_ez_0.98}. The extremums of $\left(a-b\right)$ are readily given using Eq. \ref{eq:C2} and are marked as red circles in the bottom panels of Fig. \ref{fig:potential_energy_and_libration_rotation_maps}.

The slow angle $\left(\theta-\phi\right)$ can either librate or rotate depending on the constants of motion $C_1$ and $C_2$. Examples of trajectories of both cases are plotted as equi-$C_1$ curves in the bottom panels of Fig. \ref{fig:potential_energy_and_libration_rotation_maps}. For rotations, $\cos\left(\theta-\phi\right)$ must reach both $1$ and $-1$. Using Eqs. \ref{eq:C1}-\ref{eq:C2}, we have 
\begin{equation}
    \cos\left(\theta-\phi\right)=\frac{1}{\left(a-b\right)}\left(\frac{2}{45}C_{1}-\frac{1}{12}\left(C_{2}-\left(a-b\right)^{2}\right)^{2}\right).\label{eq:cosThetaMinusPhi_vs_aMinusb}
\end{equation}
Given $C_{1}$ and $C_{2}$ the rhs. of Eq. \ref{eq:cosThetaMinusPhi_vs_aMinusb} has a global maximal value (in the $\left(a-b\right)>0$ regime) denoted $M\left(C_1, C_2\right)$. If $M\left(C_1, C_2\right)<-1$ - Eq. \ref{eq:cosThetaMinusPhi_vs_aMinusb} cannot be satisfied for any $\left(\theta-\phi\right)$ and so the pair $\left(C_1, C_2\right)$ do not represent any set of initial conditions. If $M\left(C_1, C_2\right)<1$ the slow angle $\left(\theta-\phi\right)$ is librating. If $M\left(C_1, C_2\right)>1$ both $\cos\left(\theta-\phi\right)=-1$ and $\cos\left(\theta-\phi\right)=1$ can be reached (because at $\left(a-b\right)\rightarrow\infty$ the rhs. of Eq. \ref{eq:cosThetaMinusPhi_vs_aMinusb} approaches $-\infty$) and $\left(\theta-\phi\right)$ is rotating \footnote{For analysis of Eq. \ref{eq:cosThetaMinusPhi_vs_aMinusb} we equip the reader with a link to a visualization: \url{https://www.desmos.com/calculator/wiitg5elt6}}. Since the rhs. of Eq. \ref{eq:cosThetaMinusPhi_vs_aMinusb} is monotonically increasing with $C_1$, for each $C_2$ there is therefore a minimal permitted $C_1$ and a higher minimal $C_1$ above which $\left(\theta-\phi\right)$ is rotating. The latter threshold is shown as a black dashed curve in the lower panels of Fig. \ref{fig:potential_energy_and_libration_rotation_maps}.

For the regime we solve, $e^2_z$ close to $1$, $C_K<0$ and the minimum and maximum values of the eccentricity during any such Kozai cycle are obtained at $\omega=\pm\frac{\pi}{2}$. As a result, these can be calculated using the constants $j_z$ and $C_K$ through
\begin{equation}
 3e^4_{extremum}+\left(5j^2_z-3+2C_K\right)e^2_{extremum}-2C_K=0\label{eq:polynom_for_extremal_eccentricity}.
\end{equation}
The long-term evolution of $j_z$ is obtained via Eq. \ref{eq:C4} and the evolution of $C_K$ follows 
\begin{equation}
 C_K=-\frac{3}{2}+\frac{1}{2}\alpha^{\frac{2}{3}}\left(\frac{1}{2}\left(C_{2}+C_{3}\right)-\sqrt{\frac{5}{3}}\delta\right).\label{eq:Ck}
\end{equation}
The extremal values of the eccentricity obtained from $j_z$ and $C_K$ are plotted (on a semi-log $1-e$ plot) as red dashed curves in the upper panel of Fig. \ref{fig:delta_and_jz_one_minus_e_alpha_0.01_ez_0.98}. As can be seen in the left panel, the analytical solution captures the long term evolution of the oscillations to an excellent approximation compared to the numerical integration of Eqs. \ref{eq:secular_equations}.

\section{Discussion}

In this letter we provide a concise presentation of the analysis and analytic solution for the dynamics of a test particle in a Keplerian orbit perturbed by a precessing quadrupole potential. We explicitly demonstrate the success of the solution vs. full numerical solution of the secular equations for a case that is very close to the assumptions made, i.e extremely small $\alpha$, $\beta=\beta_0$ and $e_{z}\sim1$.

Exploring the validity of the solution for wider scopes of the parameters is beyond the scope of this work, but as an example we present in the right panel of Fig. \ref{fig:delta_and_jz_one_minus_e_alpha_0.01_ez_0.98} the capability of the analytic solution to describe and approximately reconstruct the long term evolution even for higher values of $\alpha$ ($5^{\circ}$ as is numerically explored in \citep{hamers2017}) and for a value of $\beta$ slightly different from $\beta_0$.

Although the analytical model presented in this letter is directly applicable only to a small region of the parameter space (i.e test particle and small perturbation) and only for the final stages of the evolution (i.e at high eccentricity) - it serves as a basis for understanding the more complex phenomena, when the two bodies have comparable mass, and hints for the evolution farther from resonance (i.e starting with low eccentricity).

In the future, we plan to explore the validity and relevance of this model to the different astrophysical phenomena involving KLCs. In addition, we plan to relax some of the assumptions especially starting with low eccentricity or relaxing the test particle assumption. 

\begin{acknowledgments}
We thank the anonymous referee for helpful comments improving this letter. We thank Ido Barth for a useful discussion pointing the connection to coordinate moving in a potential well.
\end{acknowledgments}

\bibliography{precessing_potential_KLC}{}
\bibliographystyle{aasjournal}



\end{document}